\documentclass[aps,prl,twocolumn,nofootinbib,superscriptaddress]{revtex4}
\usepackage[hypertex]{hyperref}
\usepackage{graphicx}
\usepackage{amsmath}
\newcommand{\gsim}{\lower.7ex\hbox{$\;\stackrel{\textstyle>}{\sim}\;$}}
\newcommand{\lsim}{\lower.7ex\hbox{$\;\stackrel{\textstyle<}{\sim}\;$}}
\def\OO{{\cal O}}

\newcommand{\half}{{\frac{1}{2}  }}

\begin{document}

\pagestyle{plain}
\title{
\begin{flushright}
\mbox{\normalsize SLAC-PUB-13803}
\end{flushright}
Using Atom Interferometery to Search for New Forces}

\author{Jay G. Wacker}
\affiliation{
Theory Group, SLAC National Accelerator Laboratory, Stanford University, Menlo Park, CA 94025 
}

\date{\today}

\begin{abstract}
Atom interferometry is a rapidly advancing field and this Letter proposes an experiment based on existing technology that
can search for new short distance forces.  With current technology it is possible to improve the sensitivity by
up to a factor of $10^2$ and near-future advances will be able to rewrite the limits for forces with ranges from
100 $\mu$m to 1km.
\end{abstract}

 \maketitle

\noindent
New short distance forces are a frequent prediction of theories beyond the Standard Model and
the search for these new forces is a promising channel for discovering new physics.  Over the past 15 years
there has been rapid advances in light pulse atom interferometry (AI) and in a wide variety of settings, AI
is the most sensitive measurement.  This Letter will explore the sensitivity of AI to new forces. 
AI holds great promise in improving
currently sensitivity over a wide range of distances from roughly \mbox{100 $\mu$m} to \mbox{1 km}. 

New forces can couple to matter in many different ways; however, there is a benchmark parameterization that is frequently applicable to new forces where the potential between two particles is proportional to the mass of the
particles
\begin{eqnarray}
\label{Eq: Yukawa}
V(r) =  \alpha \frac{G_N m_1 m_2}{r} \exp(- r/\lambda)
\end{eqnarray}
where $\alpha$ is a dimensionless number that characterizes the new force's strength relative to gravity and $\lambda$ is the Compton wavelength of the particle being exchanged.   The coupling, $\alpha$, could be composition, spin or velocity dependent or have a power-law fall off rather than an exponential/Yukawa behavior; however, this parameterization is a standard benchmark and will be used in this Letter.
Theories predict a wide range of $\lambda$ and $\alpha$. Some theories give $\alpha \gsim \OO(1)$ such as gauge mediated supersymmetry theories that have moduli mediated forces \cite{ModuliForces}, large extra dimensions \cite{LED} or theories that have gravity shut off at the scale of the cosmological constant \cite{FatGravity}. Alternatively, many theories also predict $\alpha \ll 1$\cite{AdelbergerNelson}.  The most reknowned of these theories are Peccei-Quinn axions can mediate forces with  $\alpha \lsim 10^{-6}$ \cite{Axion, AxionForces}.   Thus, while it is important to continue the search for $\alpha\!\sim\!1$ to  shorter distance forces, searching for \begin{it}sub\end{it}-gravitational strength forces is also an important frontier to continue pursuing.   Finally, there are forces that are not Yukawa forces of (\ref{Eq: Yukawa}) and may intrinsically be stronger than gravity, but at long distances may show up as sub-gravitational strength forces \cite{NewForces}. See \cite{DimopoulosGeraci,Bailey} for other applications of atom interferometry to modifications of gravity.

Atom interferometry uses cold atoms that have their quantum mechanical wave packets spatially split in two and recombined.  The final interference pattern measures the phase difference between the two paths.    The experiment described in this Letter uses a source mass to create a potential that causes a relative phase between the two paths.  By subtracting off the Newtonian potential and other backgrounds, a new Yukawa potential is visible. 
The AI experiment in this Letter is effective at probing new forces in the \mbox{1mm} to \mbox{100m} range with sensitivities down to $\alpha \sim 10^{-5}$ with already proven techniques in contrast to current experimental limits that have sensitivities of $\alpha \gsim 2\!\times\!10^{-4}$\cite{AdelbergerNelson,Fischbach}.  Future improvements can increase the range of sensitivity to \mbox{100$\mu$m} to \mbox{1km} and with sensitivities down to $\alpha \sim 10^{-9}$.

\section{Atom Interferometry}
The atom interferometry used in this proposal is similar to \cite{Biedermann, ChuAI} and uses many of the techniques in \cite{StanfordAI}.
Light pulse atom interferometry uses two counter-propagating lasers that couple hyperfine degenerate ground states of alkali (or alkali-like) atoms through a near-resonance Raman transition.  
While the lasers are on, the system undergoes Rabi oscillations 
between two states having a relative momentum $\hbar k_{\text{eff}} = m v_r$.    
By performing a $\frac{\pi}{2}\!-\!\pi\!-\!\frac{\pi}{2}$ series of Raman pulses, the atom's wave packet is split into a slow and fast component, then after an interrogation time, $T$, the states are reversed and the wave packet is brought back together for the final beam splitter that interferes the two halves of the wave packet.    The maximal spatial separation of the wave packets in the interferometer is $v_r T$.   There often is an initial velocity, $v_i$, to the atom's wave packets that is used to Doppler-select the desired atomic transitions.  An initial velocity can also arise from the thermal velocities.   

Atom interferometers are always run in pairs with the same lasers driving both interferometers in order to remove laser phase noise.  The paired interferometers also reduce other common mode backgrounds such as the Sagnac effect.  The distance between these interferometers can be large and a benchmark value of several meters will be used.   Finally, interferometers are run with $N_a$ atoms simultaneously  in a bunch and the full experiment is performed $N_b$ times.  $N_a$ is determined by the rate for cooling atoms and $10^8/s$ is currently possible.  The interrogation time for the experiment will be $T\sim 0.1 s$, so a benchmark values of $N_a \sim 10^7$ is reasonable.   In several days $N_b \sim 10^7$ trials can be run and gives a shot-noise phase sensitivity of $ \delta \phi \!=\!N_a^{-\half}\! N_b^{-\half}\!\sim\!10^{-7}$. 

Atomic fountains are frequently used in AI and start with an ensemble of evaporatively cooled atoms ($\tau < \mu\text{K}$) in a magneto-optical trap (MOT) that localizes that atoms to $\sigma_x \sim 100 \mu\text{m}$.   The atoms are then launched with a velocity, $v_l$, vertically through a series of multiple Bragg or Raman transitions.   The atoms subsequently follow ballistic motion for an interrogation time, $T = g/v_l$, 
defining $v_l$ at the initial $\frac{\pi}{2}$ pules and performing the $\pi$ pulse occurs at the apex of the trajectory. 
In atomic fountains, the atoms are instantaneously in free fall and  decoupled from their environment except during
periods when they are coupled to the lasers which last  \mbox{$10^{-5}$s}.   The atoms are more isolated than  meso- or macroscopic measurements that are vibrationally coupled to the environment and essentially removes  the  ``chopping''  used to isolate torsional pendulums or cantilevers from the environment.

Configuring the interferometer as a gyroscope, with $v_l$ perpendicular to $v_r$ and $v_i$,  
allows the atom's ballistic trajectory to be parallel to a planar face of the mass.  This maximizes the time that the atoms spend close to the source mass, thus making a signal from a new short distance force as large as possible.  The interferometer is sensitive to the potential as a function of distance away from the surface of the source mass if the recoil velocity is perpendicular to this face.
By mapping out the potential, it is possible to look for new contributions beyond the Standard Model.
This configuration minimizes the Earth's gravity; however, the phase from the Corriolis force (Sagnac effect) is maximized and controlling this is an important background discussed below.

\begin{figure}[h]
\centerline{
\includegraphics[width=3.0in]{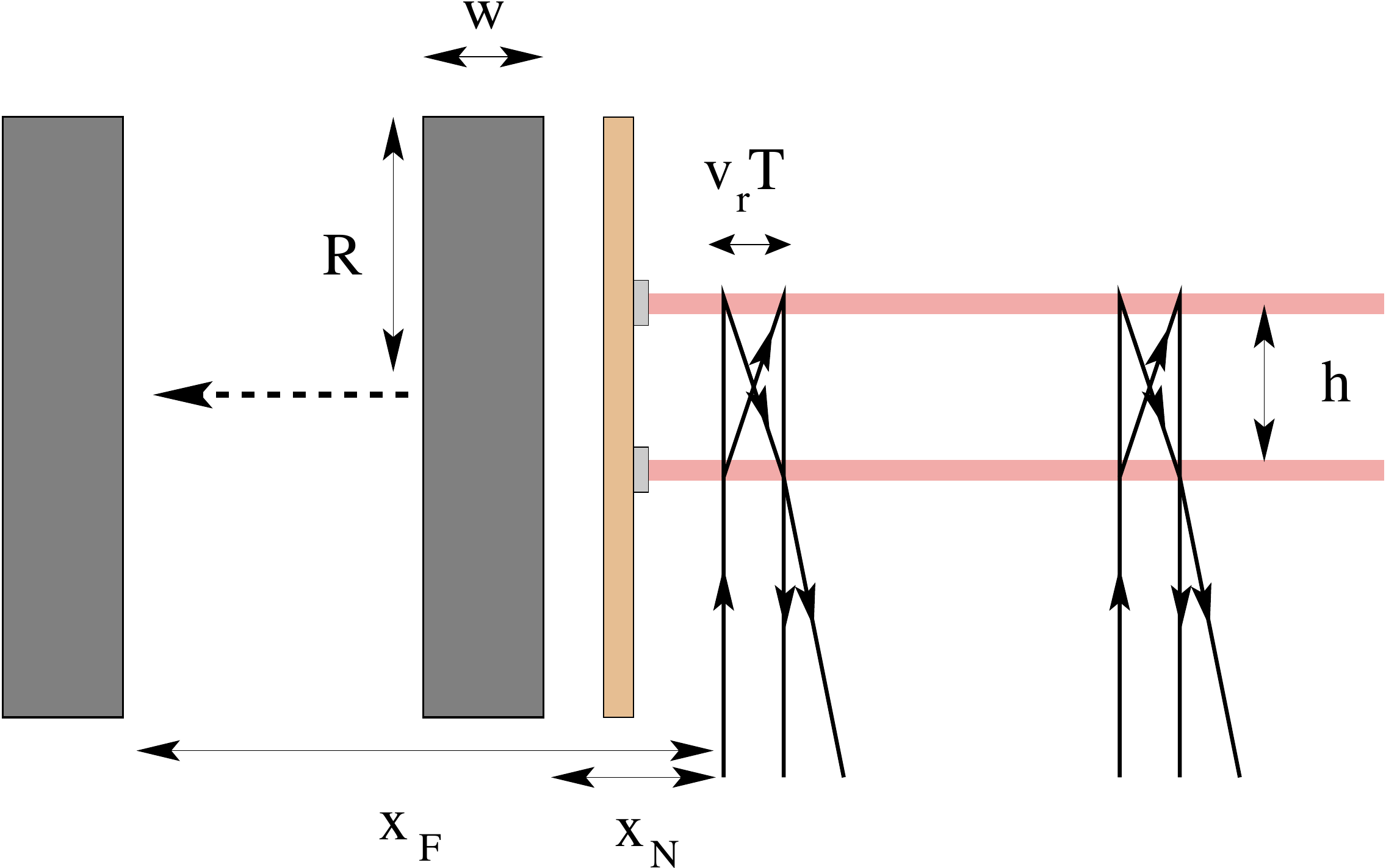}
\vspace{-0.1in}
}
\caption{
The diagram of the proposed experiment where the phase from the grey source mass is measured at distances of $x_N$ and $x_F$ from the nearer of the paired interferometers.  The plane in gold represents a thin Casimir shield fitted with retroreflectors that allow optical access for the right moving laser.}
\label{Fig: Diagram}
\end{figure}

The relative phase between the two paths arises from several different contributions and not many exact results are known.
There is a semi-classical, perturbative method for computing the phase differences and at lowest order in the potential, $V(x)$, speed of light, $c$, and the width of the wave packet, $\sigma_x$,
the general result is that the phase difference is integral of the perturbing potential over the unperturbed paths.
There are several phase difference results that will be useful in deriving the sensitivity.  The first is if the potential only depends on the position in the direction of $v_r$ and where $v_i$ is not important
\begin{eqnarray}
\label{Eq: Pot Phase}
\hbar \phi &=& ( V(v_r T) - V(0) ) T\\
\label{Eq: Acc Phase}
& =&  \frac{V'}{m} \hbar k_{\text{eff}} T^2 +  \frac{V''}{m^2} (\hbar k_{\text{eff}})^2 T^3 + \cdots
\end{eqnarray}
where the initial position of the atom  is taken to be $x=0$.  In the second half the potential has been Taylor expanded, if applicable.    The first term is proportional to the acceleration, and the second term 
is referred to as the recoil phase because it vanishes as $m$ becomes large, keeping $k_{\text{eff}}$  fixed.   
Finally, if $v_i$ is important, the above expression becomes
\begin{eqnarray}
\label{Eq: Phase vi}
\hbar \phi =  \left(\frac{ W(x_1+x_2) - W(x_2) - W(x_1) + W(0)}{\big(x^{-1}_1- x_2^{-1}\big)^{-1}}\right)T
\end{eqnarray}
where $x_1= (v_r\hspace{-0.015in}+\hspace{-0.015in}v_i)T$, $x_2=v_iT$, and  $W(x) = \int^x\!\!\! dx' V(x')$.   
If the potential depends on the distance in the $v_l$ direction, no closed form is possible, but simple expressions can be obtained if  $V(x,y)$ is Taylor expandable.

\section{ Proposed Experimental Design}
Newton's constant is not known to a precision better than $10^{-4}$, so it is necessary to perform a series of measurements to remove absolute sensitivity to  $G_N$.  The most straightforward manner to remove the uncertainty in $G_N$ is to test the $r^{-1}$ behavior of gravity by having a moveable source mass.   The source mass will be taken to be a cylinder of radius, $R$, and width, $w$, and density, $\rho$, with the  circular face of the cylinder forming a vertical plane.  This geometry is motivated by calculational simplicity and none of the results depend sensitively upon the geometry.  The height of the atomic fountains, $h= v_l^2/2g < R$, will take place near the center of the cylinder's face. 

The Newtonian potential  near the center of the cylinder can be calculated in the far field, $x \gg R$, and near field, $x \ll R$,  limits near the center of the cylinder.  In the far field limit
\begin{eqnarray}
\label{Eq: Newton 1}
\frac{V_N(x,r)}{G_N \rho m} =  \frac{ \pi R^2 w}{x}\left( 1 - \frac{w}{x} - \frac{3 R^2}{16 x^2} - \frac{r^2}{2x^2}\right)
\end{eqnarray}
where $x$ is the distance to cylinder's face and $r$ is the distance from the cylinder's center.\footnote{The absence of an $\OO(r/x)$ correction to the potential is due to the cylindrical symmetry and will not be important for the results.}  
In the near field limit the Newtonian potential is given by
\begin{eqnarray}
\label{Eq: Newton 2}
\nonumber
\frac{V_N(x,r)}{G_N\rho m} &=&  \frac{V_0(r)}{m} + 2\pi x\left( (w^2\!+\!R^2)^\half\!-\!R\!-\!w\right)\\
&& \hspace{-0.2in} + \frac{\pi w x^2}{(w^2\!+\!R^2)^\half}  - \frac{ \pi r^2 x}{2}\!\!\left( \frac{R^2}{(w^2\!+\!R^2)^{\frac{3}{2}} }\!-\!\frac{1}{R}\right)
\end{eqnarray}
where  $V_0(r)$ is an unmeasurable function, independent of the distance from the source.

There are several limits necessary for the Yukawa potential.   The first is $\lambda \ll R, w$ where the potential from cylinder is
\begin{eqnarray}
\frac{V_Y(x)}{\alpha G_N\rho m} =\begin{cases}
 2 \pi \lambda^2 \exp(-x/\lambda)  &\lambda <  R^2/x\\
  \pi \lambda R^2x^{-1} \exp(-x/\lambda)& \lambda > R^2/x\end{cases}  .
\end{eqnarray}
For $\lambda \gg R, w$, the Yukawa potential looks Newtonian plus a correction arising from Taylor expanding the exponential.
In the far field limit ($x\gg R$) the Yukawa potential is
\begin{eqnarray}
\frac{V_Y(x)}{\alpha G_N\rho m} = \frac{V_N(x)}{G_N\rho m} + \frac{ \pi R^2 w x}{ 3 \lambda^2}.
\end{eqnarray}
Notice the absence of an $\OO(\lambda^{\text{-}1})$ term because it is always an unmeasurable constant.   In the near field limit ($x\ll R$),
the Yukawa potential becomes
\begin{eqnarray}
\frac{V_Y(x)}{\alpha G_N\rho m} = \frac{V_N(x)}{G_N\rho m} + \frac{ 2 \pi x}{3\lambda^2} \left( (w^2\!+\!R^2)^{\frac{3}{2}} - R^3 -w^3\right).
\end{eqnarray}

The physical size of the experiment ultimately limit the sensitivity.  The size of the source mass sets the $\lambda$ with maximum sensitivity and the benchmark value used is $R \le 1 \text{ m}$ and $w < 1 \text{ m}$.  The other relevant physical constraint is how near the source mass can get to the interferometers, $x^{\text{min}}_N \ge 200 \mu\text{m}$.  The distance that the  source mass can be moved from the interferometers does not limit the sensitivity so long as $x^{\text{max}}_F \gsim R,w$. 

The strategy to distinguish Newtonian gravity from a new Yukawa potential is to perform a near  and a far measurement of the phase.  The near measurement fixes what the Newtonian prediction is for the far measurement.   
If the inclusion of a Yukawa potential with strength $\alpha$ and range $\lambda$ causes a difference between the far prediction and the far measurement greater than the shot noise limit, then this Yukawa potential is visible assuming that no backgrounds or uncertainties are larger.
The scaling of the limits depend on $\lambda$ and there two cases to be considered.  
 If, $\lambda\!<\!R$, it is possible to get within the $r^{-1}$ behavior of the Yukawa potential and move outside its range; furthermore, the radius of the cylinder is not determining the sensitivity. 
In the second case $\lambda \!>\! R,w$, it is possible to get within the $r^{-1}$ and outside the range of the potential, but the size of the source mass is determining the sensitivity.  For large $\lambda$, it may seem beneficial to move the source mass further from the source to make the difference between a Yukawa potential and the Newtonian potential more pronounced, growing as $r^2/\lambda^2$; however, the size of the Newtonian phase shift is falling as $r^{\text{-}2}$ meaning that there is no parametric gain for moving the source mass a distance greater than $R$ from the interferometers for any value of $\lambda$.

The biggest challenge of this experiment is the strategy of using a near measurement to extrapolate to a far prediction.  The challenge is knowing both the source mass position and orientation precisely enough to make a $10^{-6}$ or better prediction for the far Newtonian prediction.   Examining the subleading terms in the Newtonian potentials in (\ref{Eq: Newton 1}) and (\ref{Eq: Newton 2}), if there is an uncertainty in the position of the cylinder of $\delta x$, then there is an uncertainty in the Newtonian potential of $\delta x/ R$ or $\delta x/x$, respectively.   By having $h<R$ reduces the uncertainty in the Newtonian prediction from uncertainty in the initial height of the atom.   It is challenging but achievable to know the position of the block to $\delta x\sim 1\mu\text{m}$ to be sensitive to $\alpha \sim 10^{-6}$; however, future improvements in sensitivity will not be limited by this.   A better solution to knowing the position and orientation of the block is to use multiple interferometers situated near $r \sim R$ to actively measure the position and orientation of the source.    
Since the edge effects become $\OO(1)$ near the edges, it is possible to use these additional measurements to locate the block and then use a central interferometer to use the near measurement to make the far prediction.   This strategy will require six additional interferometers to over-determine the source's solid body coordinates.   The remaining challenge is to keep these seven interferometers locked in place as the source mass moves, but this should be possible to a much greater accuracy.  There is a residual uncertainty coming from ``jitter'' in the location of the MOTs.   This motion is of the order of $\delta x \sim 1 \mu\text{m}$; however, it is stochastic with respect to the number of bunches of atoms, $N_b$, and therefore the uncertainty is  $\delta x/RN_b^\half\sim 10^{-9}$ and only limiting after several rounds of improvements.

The Sagnac effect, the phase induced by the Corriolis force is large, $ \phi_{\text{Sag}} = k_{\text{eff}} v_l \omega_{\oplus}T^2\sim 10^2$,
where $\omega_\oplus$ is the angular velocity of the earth.  The use of paired interferometers described above cancels the leading order Sagnac effect.  The novel method of actively reducing the Sagnac effect is to rotate the lasers to compensate for the rotation of the Earth.  It is possible to rotate lasers with nanoradian precision which reduces the Sagnac effect by a factor of $10^{-4}$.   The dominant way that the Sagnac effect contaminates the signal is from variation in relative  $v_l$ of the two interferometers
which is stochastic in the number of bunches, reducing the size by $N_b^{-\half}$.  The final reduction in Sagnac contamination must come from vibration isolation and requires $ \delta v^{\text{rel}}_l/v_l < 10^{-2}$. 

The Casimir potential between an atom and a conducting plate is given by $V(x)= \hbar c \alpha_0/x^4$, with $\alpha_0$ being the polarizability of the atom and is $59.4 \mbox{\AA}{}^3$ for Cs\cite{Polarizability}.  The phase difference arising from the Casimir potential  arises from (\ref{Eq: Pot Phase}) or (\ref{Eq: Phase vi}) rather than (\ref{Eq: Acc Phase}) because of  its rapid fall off.  These reduce the size of the Casimir force relative to a macroscopic measurement device and it is not important for distances,  (and consequently $\lambda$) greater than  $\OO(100 \mu\text{m})$, but becomes important at shorter distances.
In order to gain to reduce any residual phases, having a thin Casimir shield ($\OO(50\mu\text{m})$ in width) that isolates the atom from the source mass' position is helpful.   While the the Casimir potential from the shield will still be measurable, it will not be correlated with the mass' position and therefore not directly a systematic effect.   The primary way the Casimir potential enters as a background is by having the source mass  deflect the shield and additionally through the ``jitter'' of the MOTs. 
The deflection of the shield is sufficiently small so long as the source mass is $\OO(30 \mu\text{m})$ from the shield.   The jitter  gives a phase uncertainty of $\delta \phi_{\text{Cas}}/\phi_{\text{Cas}}\! \sim\! (\delta x/x) /N_b^\half\!\sim\! 10^{\text{-}4}$ for $x= 100 \mu\text{m}$ and is sufficiently small.\footnote{The calculations for the phase shifts use the semiclassical approximation and break down when $x\sim \sigma_x \sim 100\mu\text{m}$.}    In addition to isolating the atoms, it will be necessary to attach retroreflectors on the shield to provide optical access for the outward propagating lasers necessary to drive the Raman transitions.

\section{Expected Sensitivity and Future Improvements}
\begin{figure}[h]
\centerline{
\includegraphics[width=3.0in]{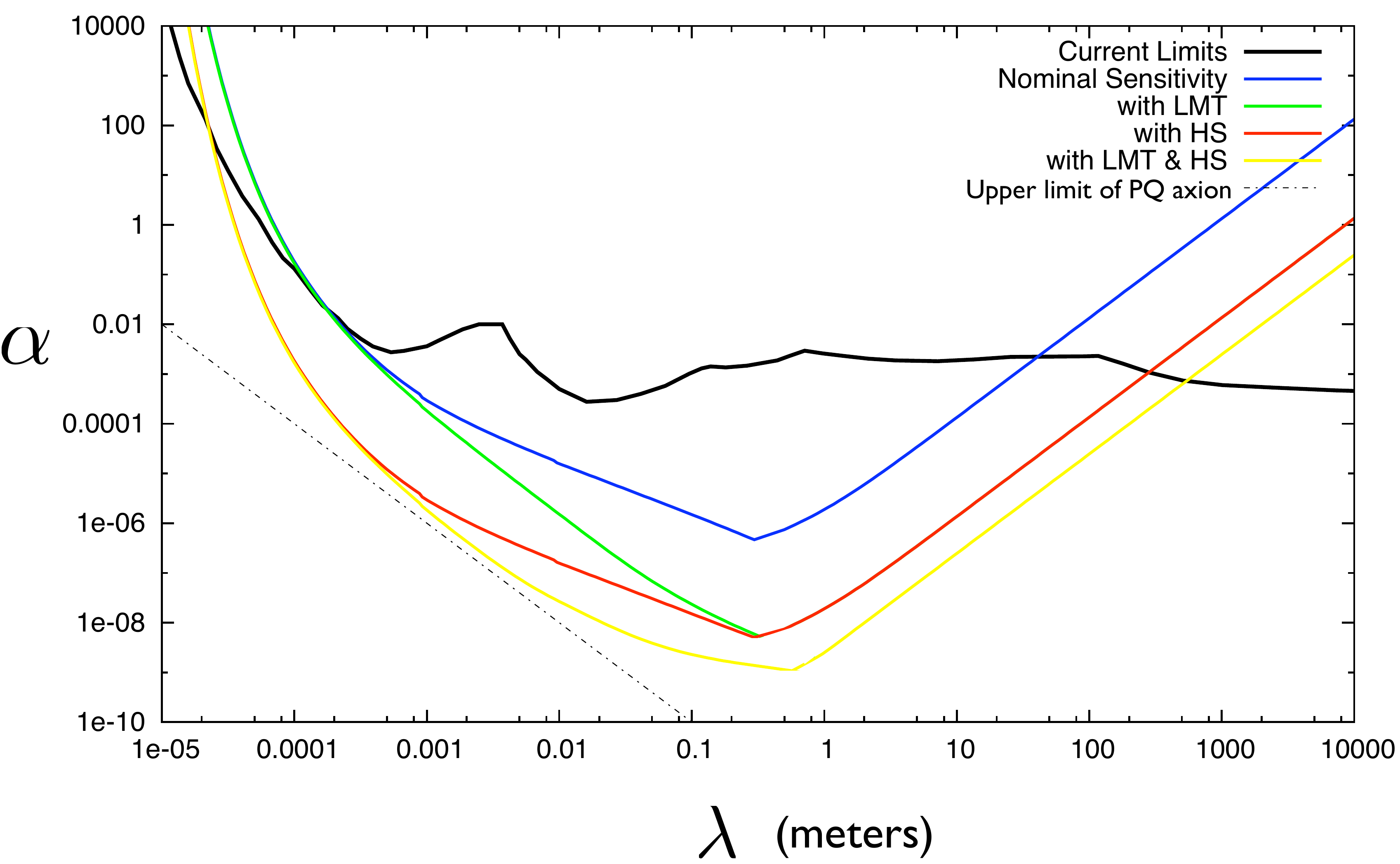}
\vspace{-0.1in}
}
\caption{
The current limits on new forces and where AI can probe with current nominal sensitivity
and future improvements described in the text.
}
\label{Fig: Sensitivity}
\end{figure}

With the results of the previous section, the sensitivity to $\alpha$ and $\lambda$ can be computed.  
Using the previously stated benchmarks for Cs the phase sensitivity of $\delta \phi =10^{-7}$ from $N_aN_b=10^{14}$,
$R,w = 1 \text{m}$, $x_{\text{N}}^{\text{min}}= 200 \mu\text{m}$, $h=10 \text{cm}$, $\rho = 12 \text{g/cm}^3$ for Pb
 and $\tau = 10\text{nK}$, the sensitivity curve is plotted in Fig. \ref{Fig: Sensitivity} relative in blue to current experimental limits shown black.  The peak sensitivity is $\alpha \sim 3\!\times\!10^{-6}$ and occurs for $\lambda \sim 0.5 \text{m}$ which is determined by the size of the source mass.   For $1 \text{mm} < \lambda < 10 \text{m}$, current AI is more sensitive than existing experiments.  Additionally, from $100 \mu\text{m} < \lambda < 1\text{mm}$ current AI techniques matches the experimental limits.    This motivates considering future experimental improvements to if sensitivity is possible down to $\lambda \sim 100\mu\text{m}$.

In the near future it will be possible to increase the magnitude of the momentum transferred to the atoms in the interferometer using a technique called large momentum transfer (LMT).  LMT uses repeated Raman or Bragg transitions during the laser pulses that impart on the atoms up to 100 times more momentum as a single Raman transition.  LMT  significantly impacts sensitivity for larger $\lambda$, increasing sensitivity by a factor of 100; however, for shorter wavelengths, where $\lambda < v_r T$,  there is no significant gain to sensitivity because the atoms immediately leave the region being sourced by Yukawa potential.\footnote{If there is an initial velocity, $v_i$, from either a need to Doppler-select atomic transitions or thermal motion, there is a small gain at short distances proportional to $(1+v_i/v_r)^{\text{-}1}$ shown in (\ref{Eq: Phase vi}).}  The sensitivity is shown in Fig. \ref{Fig: Sensitivity} in green.

On the short distance front, increases in sensitivity must come through increases in phase sensitivity. One possibility is to increase the number of atoms being used in the experiment, which is limited by the length of the experiment and the rate for cooling atoms.   In the future, a more effective way to increase sensitivity would be to use wave packets of entangled atoms where it is possible to have the sensitivity scale as $N_a^{-1}$  (in contrast to $N_a^{-\half}$) which is known as Heisenberg limited statistics (HS).    HS potentially allows significant gains in sensitivity and the sensitivity curve is shown for a phase sensitivity of $\delta \phi = 10^{-9}$  in red in Fig. \ref{Fig: Sensitivity}.   In yellow, the sensitivity is shown for combining a LMT factor 100 and $\delta \phi = 10^{-9}$; however, at this level the stochastic uncertainty in the Newtonian prediction starts to become limiting.    There are some caveats that may limit the sensitivity at shorter distances.   The first is that all of the phase results  are computed in the semi-classical limit; however, when $\lambda < \sigma_x$ this approximation breaks down and a more complete quantum mechanical treatment is necessary.   Additionally,  the interaction between the Casimir shield and atoms must be considered in much more detail, particularly for the case of HS where preventing the decoherence of the wave packet is  necessary.

This Letter has demonstrated that atom interferometry has the potential to improve the sensitivity to new forces in the  $100 \mu\text{m}$ to $1 \text{km}$ range using current technology or technologies that will be available in the near future.  Additionally, atom interferometry can be useful in testing other types of forces other than the standard $\alpha\text{-}\lambda$ Yukawa potential such as spin, velocity and composition dependent forces.

\begin{acknowledgments}
JW would like to thank M. Kasevich, P. Graham and S. Dimopoulos for collaboration during early stages of this work and also G. Biederman, B. Dobrescu, and J. Hogan for illuminating conversations.  JW is  supported by the DOE under contract DE-AC03-76SF00515.   JW is supported by the DOE Outstanding Junior Investigator Award. 
\end{acknowledgments}


\begin{thebibliography}{99}

\bibitem{AdelbergerNelson}
  E.~G.~Adelberger, B.~R.~Heckel and A.~E.~Nelson,
  Ann.\ Rev.\ Nucl.\ Part.\ Sci.\  {\bf 53}, 77 (2003)
  [arXiv:hep-ph/0307284].

\bibitem{LED}
N.~Arkani-Hamed, S.~Dimopoulos and G.~R.~Dvali,
Phys.\ Rev.\ D {\bf 59}, 086004 (1999) [arXiv:hep-ph/9807344].
I.~Antoniadis, N.~Arkani-Hamed, S.~Dimopoulos and G.~R.~Dvali,
Phys.\ Lett.\ B {\bf 436}, 257 (1998) [arXiv:hep-ph/9804398].
N.~Arkani-Hamed, S.~Dimopoulos and G.~R.~Dvali,
Phys.\ Lett.\ B {\bf 429}, 263 (1998) [arXiv:hep-ph/9803315].


\bibitem{ModuliForces}
S.~Dimopoulos and G.~F.~Giudice,
``Macroscopic Forces from Supersymmetry,''
Phys.\ Lett.\ B {\bf 379}, 105 (1996) [arXiv:hep-ph/9602350].

\bibitem{FatGravity}
  R.~Sundrum,
  Nucl.\ Phys.\ B {\bf 690}, 302 (2004)   [arXiv:hep-th/0310251].


\bibitem{Axion}
R.~D.~Peccei and H.~R.~Quinn,
Phys.\ Rev.\ Lett.\  {\bf 38}, 1440 (1977).
F.~Wilczek,
Phys.\ Rev.\ Lett.\  {\bf 40}, 279 (1978).
M.~Dine, W.~Fischler and M.~Srednicki,
Phys.\ Lett.\ B {\bf 104}, 199 (1981).
M.~A.~Shifman, A.~I.~Vainshtein and V.~I.~Zakharov,
Nucl.\ Phys.\ B {\bf 166}, 493 (1980).
J.~E.~Kim,
Phys.\ Rev.\ Lett.\  {\bf 43}, 103 (1979).

\bibitem{AxionForces}
J.~E.~Moody and F.~Wilczek,
Phys.\ Rev.\ D {\bf 30}, 130 (1984).
R.~Barbieri, A.~Romanino and A.~Strumia,
Phys.\ Lett.\ B {\bf 387}, 310 (1996)
[arXiv:hep-ph/9605368].
M.~Pospelov,
Phys.\ Rev.\ D {\bf 58}, 097703 (1998)
[arXiv:hep-ph/9707431].

\bibitem{NewForces}
  B.~A.~Dobrescu and I.~Mocioiu,
  JHEP {\bf 0611}, 005 (2006)
  [arXiv:hep-ph/0605342].
 
 \bibitem{DimopoulosGeraci}
  S.~Dimopoulos and A.~A.~Geraci,
  Phys.\ Rev.\  D {\bf 68}, 124021 (2003)
  [arXiv:hep-ph/0306168].
\bibitem{Bailey}
  Q.~G.~Bailey and V.~A.~Kostelecky,
  Phys.\ Rev.\  D {\bf 74}, 045001 (2006)
  [arXiv:gr-qc/0603030].


 
 
\bibitem{Fischbach}
E.~Fischbach and C.L.~Talmadge. {\em The Search for Non-Newtonian Gravity.} New York: Springer-Verlag (1999).
 
 
\bibitem{Biedermann}
G.~Biedermann,
``Gravity Tests, Differential Accelerometry and Interleaved Clocks with Cold Atom Interferometers,''
(2008).
\bibitem{ChuAI}
  K.~Y.~Chung, S.~w.~Chiow, S.~Herrmann, S.~Chu and H.~Muller,
  Phys.\ Rev.\  D {\bf 80}, 016002 (2009)
  [arXiv:0905.1929 [gr-qc]].
  H.~Muller, S.~w.~Chiow, S.~Herrmann, S.~Chu and K.~Y.~Chung,
  Phys.\ Rev.\ Lett.\  {\bf 100}, 031101 (2008)
  [arXiv:0710.3768 [gr-qc]].


\bibitem{StanfordAI}
S.~Dimopoulos, P.~W.~Graham, J.~M.~Hogan and M.~A.~Kasevich,
arXiv:0802.4098 [hep-ph].
  S.~Dimopoulos, P.~W.~Graham, J.~M.~Hogan and M.~A.~Kasevich,
  Phys.\ Rev.\ Lett.\  {\bf 98}, 111102 (2007)
  [arXiv:gr-qc/0610047].
  A.~Arvanitaki, S.~Dimopoulos, A.~A.~Geraci, J.~Hogan and M.~Kasevich,
  Phys.\ Rev.\ Lett.\  {\bf 100}, 120407 (2008)
  [arXiv:0711.4636 [hep-ph]].
  S.~Dimopoulos, P.~W.~Graham, J.~M.~Hogan, M.~A.~Kasevich and S.~Rajendran,
  Phys.\ Rev.\  D {\bf 78}, 122002 (2008)
  [arXiv:0806.2125 [gr-qc]].
  S.~Dimopoulos, P.~W.~Graham, J.~M.~Hogan, M.~A.~Kasevich and S.~Rajendran,
  Phys.\ Lett.\  B {\bf 678}, 37 (2009)
  [arXiv:0712.1250 [gr-qc]].
  \bibitem{Berman}
P.R.~Berman, Ed. {\em Atom Interferometry\/} (New York: Acad. Pr., 1997).

\bibitem{Fixler}
J. B. Fixler, {\it et al.},  Science {\bf 315}, 74 (2007).


\bibitem{PhysRevLett.78.2046}
T.~L. Gustavson, P.~Bouyer, and M.~A. Kasevich.
\newblock Phys. Rev. Lett. {\bf 78}, 2046 (1997).

\bibitem{PhysRevLett.81.971}
M.~J. Snadden, {\em et~al.\/}
\newblock Phys. Rev. Lett. {\bf 81}, 971 (1998).

\bibitem{0026-1394-38-1-4}
A.~Peters, K.~Y. Chung, and S.~Chu.
\newblock Metrologia {\bf 38}, 25 (2001).

\bibitem{Marion}
H. Marion, et. al., Phys. Rev. Lett. {\bf 90}, 150801 (2003).

\bibitem{Bertoldi}
A. Bertoldi, {\em et~al.\/}, Eur. Phys. J.D, {\bf 271} (2006).

\bibitem{atomicsources}
C. Pethick and H. Smith, {\it Bose-Einstein Condensation in Dilute Gases} (New York: Cambridge Univ. Pr., 2002).

\bibitem{KasevichChu}
M.~Kasevich and S.~Chu.
\newblock Phys. Rev. Lett. {\bf 67}, 181 (1991).

\bibitem{LMT}
J.~H. Denschlag, {\em et~al.\/}.
J. Phys. B: At. Mol. Opt. Phys. {\bf 35}, 3095 (2002).
J.M.~McGuirk,  {\em et~al.\/} Phys. Rev. Lett. {\bf 85} 4498 (2000).
M.~Weitz,  {\em et~al.\/} Phys. Rev. Lett. {\bf 73} 2563 (1994).

\bibitem{HS}
D. Leibfried, et. al. Science {\bf 304}, 1476.
A. K. Tuchman,  {\em et~al.\/}  Phys. Rev. A. {\bf 74}, 053821 (2006).



\bibitem{Polarizability}
J. M. Amini and H. Gould,
Phys.\ Rev.\ Lett. {\bf 91}, 153001 (2003).


\end{thebibliography}
\end{document}